\documentclass[conference]{IEEEtran}
\IEEEoverridecommandlockouts
\usepackage{cite}
\usepackage{amsmath,amssymb,amsfonts}
\usepackage{algorithmic}
\usepackage{graphicx}
\usepackage{textcomp}
\usepackage{xcolor}
\usepackage{hyperref}
\usepackage{soul}
\usepackage{booktabs}
\usepackage{fancyhdr}
\usepackage{flushend}
\definecolor{paul}{RGB}{0,0,0}
\definecolor{paul_2}{RGB}{0,0,0}
\definecolor{tre}{RGB}{0,0,0}
\definecolor{sadaf}{RGB}{0,0,0}
\definecolor{rich}{RGB}{0,0,0}
\definecolor{maybe}{RGB}{0,0,0}
\def\BibTeX{{\rm B\kern-.05em{\sc i\kern-.025em b}\kern-.08em
    T\kern-.1667em\lower.7ex\hbox{E}\kern-.125emX}}
    
\begin{document}

\title{\textcolor{sadaf}{Placing the Control Plane: Cloud-Native Federated Learning Across Batch-Scheduled HPC}}

\author{
    \IEEEauthorblockN{
        Paul J. Wright \IEEEauthorrefmark{1},
        Huw Day \IEEEauthorrefmark{2},
        Ryan K. Daniels \IEEEauthorrefmark{3}, 
        Simon Li \IEEEauthorrefmark{4}, 
        Ruairidh MacLeod \IEEEauthorrefmark{5}, \\
        Rob Baxter \IEEEauthorrefmark{4},
        Abdelwahab Kawafi \IEEEauthorrefmark{1},
        Richard Lane \IEEEauthorrefmark{1},
        Christopher Woods \IEEEauthorrefmark{1},
        and
        Sadaf R. Alam \IEEEauthorrefmark{1}
        } \\
    \IEEEauthorblockA{
        \IEEEauthorrefmark{1}\textit{Bristol Centre for Supercomputing}, \textit{University of Bristol}, Bristol, United Kingdom
        }
    \IEEEauthorblockA{
        \IEEEauthorrefmark{2}\textit{School of Engineering Mathematics \& Technology}, \textit{University of Bristol}, Bristol, United Kingdom}
    \IEEEauthorblockA{
        \IEEEauthorrefmark{3}\textit{Research Computing Services}, \textit{University of Cambridge},
        Cambridge, United Kingdom
    }
    \IEEEauthorblockA{
        \IEEEauthorrefmark{4}\textit{DARE UK}, \textit{Health Data Research UK},
        London, United Kingdom
    }
    \IEEEauthorblockA{
        \IEEEauthorrefmark{5}\textit{EPCC}, \textit{University of Edinburgh},
        Edinburgh, United Kingdom
    }
    \thanks{Corresponding author: Paul J. Wright (email: paul.wright@bristol.ac.uk)}
}

\maketitle
\thispagestyle{fancy}
\lhead{}
\rhead{}
\chead{}
\lfoot{\footnotesize{
SC26 Workshops, November 15-20, 2026, Chicago, Illinois, USA
\newline 979-8-3195-1221-5/26/\$31.00 \copyright 2026 IEEE}}
\rfoot{}
\cfoot{}
\renewcommand{\headrulewidth}{0pt}
\renewcommand{\footrulewidth}{0pt}

\begin{abstract}
The United Kingdom's AI Research Resource (AIRR) opens national supercomputers to a community familiar with always-reachable cloud services. With batch-scheduled HPC, a user requests a time-bounded allocation that is then reclaimed. Federated Learning (FL) exposes this tension, pairing a long-lived control plane with short-lived training. We characterize the resulting placement asymmetry: the coordination server sees only model updates and control metadata, so it can sit on neutral infrastructure, whereas the client control process is bound to the participating site's data and must remain within its trust boundary. The outbound-only connection this implies is also how edge clients connect, though the batch-scheduler mismatch is specific to HPC. We demonstrate a cross-silo federation spanning three independently administered systems (Isambard-AI, Isambard 3, and Zenith) using unmodified production FL frameworks (FLARE-wrapped Flower), measure the cost the client placement incurs, and propose Kubernetes-as-a-service as the in-boundary home for the persistent process.

\end{abstract}

\begin{IEEEkeywords}
Federated Learning, Cloud Computing, Edge Computing, High-Performance Computing
\end{IEEEkeywords}

\section{Introduction}
\label{sec:intro}

The United Kingdom's AI Research Resource (AIRR) opens national supercomputing to a broader community than these systems traditionally served. Delivered through Isambard-AI~\cite{mcintoshsmith2024isambardaileadershipclasssupercomputer} at the Bristol Centre for Supercomputing (BriCS) and Dawn and Zenith at the University of Cambridge, AIRR is open to academia, the public sector, and industry, including Small and Medium Enterprises (SMEs) and startups, some supported by the government's Sovereign AI program. These users arrive at high-performance computing (HPC) systems, where a batch scheduler (e.g., Slurm~\cite{10.1007/10968987_3}) queues jobs and allocates bounded blocks of node-hours. A user requests a time-bounded resource allocation; the job waits its turn, runs, and then its resources are reclaimed.

While this model enables national supercomputers to function as shared resources, it differs from the typical cloud allocation model with which many users are familiar, where services remain reachable until released and environment lifetimes are not bounded by a scheduler. This tension is structural and recurs wherever a persistent service must operate within a bounded allocation.

Supporting such workloads requires the coexistence of both models, a practice that is increasingly adopted at BriCS. In addition to batch-scheduled HPC systems, \textcolor{sadaf}{we operate various cloud-hosted services\cite{alam2024}} and are developing a Kubernetes-as-a-service platform secured with multi-tenant RDMA (Remote Direct Memory Access) to the Slingshot interconnect~\cite{11186471, watson2026multitenant},  
providing a foundation for governed services, e.g., Trusted Research Environments (TREs)~\cite{hdra2021,satre2026}. 

Federated learning (FL)~\cite{ZHANG2021106775, kairouz2021advancesopenproblemsfederated}, the workload examined in this paper, is a particularly revealing example because it combines long-lived coordination services with scheduler-managed computation within a single application. Conventional machine learning assumes data can be collected in a single location and attached to the compute resource before training begins. Clinical records and many other sensitive research datasets, however, are \textcolor{sadaf}{often} tied to a governance boundary an institution establishes around them---what we call, throughout, a \emph{site}---by regulations, contracts, or sovereignty requirements that prevent such centralization~\cite{Rieke2020, Sheller2020}.

\textcolor{rich}{FL is often studied at the device edge, across many low-powered devices~\cite{mcmahan2017communication, hard2018federated}. Its defining constraint, however, is data locality rather than device count or scale~\cite{Li_2020}, a constraint that persists when the paradigm is inverted, with a few powerful HPC nodes instead.}

\textcolor{sadaf}{Rather than bringing data to the model, FL trains a model using decentralized data. The \emph{coordination server} starts a typical star-shaped FL round by broadcasting the current global model weights to a set of \emph{clients}. Each client trains the model for several local epochs on only the data it can access. The coordination server collects these updates in the form of weights or gradients, and updates the global model (e.g., using the Federated Averaging algorithm~\cite{mcmahan2017communication}), finishing a round. The updated global model is then distributed, and the process repeats for multiple rounds.}

This execution model separates naturally into a long-lived control plane and short-lived, compute-intensive training, and the two interact with batch scheduling in fundamentally different ways. \textcolor{black}{The control plane has two kinds of member: one \emph{coordination server} for the federation, and, for each client (the participant at a site), a control process (the \emph{client control process}) that holds credentials and drives training.} The coordination server must remain reachable throughout the federation's lifetime, from the first to the last round, \emph{a condition that no single bounded allocation can guarantee}.
\textcolor{black}{The client control process, by contrast, admits two deployment regimes, either holding a training allocation for the federation's lifetime, or handing compute back to the scheduler between rounds.} 
In both \textcolor{paul}{regimes}, the core issue is the need for a persistent process that the batch environment cannot host. 
The two control-plane members differ in degree of exposure. 
The coordination server sees only model updates and control metadata (it is not information-free, but far less exposing than raw data) and can sit on \emph{neutral infrastructure} that every site agrees to trust with that residual exposure. 
In contrast, the client control process, bound to the data, admits no such exit and must remain within the site's trust boundary. 
Privacy-preserving mechanisms 
reduce what the coordination server can infer, \textcolor{sadaf}{and so widen the range of conditions under which off-site placement is acceptable.}
This placement asymmetry defines the architectural constraint explored throughout this paper.

We examine these challenges across three independently administered systems: Isambard-AI and Isambard 3, operated by BriCS, each with its own network, scheduler, and administrative boundary, and Zenith at the University of Cambridge, with the federation crossing both system and organizational boundaries. In our case, each hosts a single participating site, so in this deployment, site and system coincide. We present two primary contributions, focusing on deployment: 

\begin{enumerate}
    \item{
    \textcolor{sadaf}{We characterize where persistent FL control processes can live under the constraints of batch scheduling, network posture, and trust boundaries.}
    The coordination and client tiers face this placement problem differently, and we \textcolor{sadaf}{frame} the difference as a placement asymmetry: the 
    coordination tier can leave the site boundary; the data-adjacent client tier cannot. The characterization is grounded in a running deployment across three independently administered systems.}
    \item{
    \textcolor{paul}{We propose a placement design that resolves the mismatch on the infrastructure side and is independent of how training jobs are launched: scheduler integrations and scheduling-aware algorithms change how the batch tier is driven, not where the persistent tiers must live. We deploy production frameworks with no HPC-specific modifications, demonstrate the coordination tier on neutral, always-on infrastructure, and propose a path to a fully in-boundary client via Kubernetes-as-a-service.}}
\end{enumerate}

\section{Background}
\label{sec:background}

FL is a workload in the edge--cloud--HPC continuum. It began with cross-device training at the edge~\cite{hard2018federated}, and has since expanded to cross-silo deployments across cloud and national HPC resources. A single federation may include all of these. Several FL frameworks target this continuum. Production runtimes such as FLARE~\cite{roth2022nvflare,roth2024empoweringfederatedlearningmassive}, and Flower~\cite{beutel2022flower} are designed for cloud and enterprise settings, while at the HPC end of the continuum, APPFL~\cite{ryu2022appfl, li2024advances} 
offers a privacy-preserving framework that scales across heterogeneous research computing environments, and OmniFed
~\cite{tyagi2025omnifedmodularframeworkconfigurable} targets configurable deployment from the edge to HPC.

Earlier work on bringing FL to HPC has followed two routes. The first adapts the algorithm to tolerate scheduling variability, while the second pairs an HPC-focused framework with orchestration middleware. This includes StreamFlow through Cross-Facility Federated Learning (xFFL~\cite{COLONNELLI20243}
) and APPFL with Globus Compute across four Department of Energy supercomputers~\cite{li2026scalablecrossfacilityfederatedlearning}, whose authors note that existing cross-silo frameworks, including FLARE, are ``primarily designed for cloud and enterprise settings and do not address HPC-specific constraints.'' Common to both is the premise that the batch environment is fixed, and that it is the federation---its algorithm or its deployment machinery---that must adapt.

We take a different route, in a direction also explored by Oak Ridge National Laboratory at NVIDIA FLARE Day 2024\footnote{\url{https://nvidia.github.io/NVFlare/flareDay/}}. Rather than adapting production FL to the scheduler via HPC-specific middleware, 
\textcolor{sadaf}{we investigate running production FL on HPC with no HPC-specific modifications, resolving the mismatch on the infrastructure side rather than in the framework or its scheduling logic}. 

\textcolor{black}{This ``zero re-engineering'' approach shifts the burden of HPC integration from the researcher to the facility. Rather than stripping out native networking logic and introducing scheduler-aware middleware to move a workflow onto a cluster, practitioners deploy unmodified codebases, and the cloud-native infrastructure tier absorbs the mismatch, preserving each framework's native behaviors, tooling, and out-of-the-box security posture.}

\textcolor{rich}{We therefore concentrate on FLARE and Flower, two open-source production frameworks that both support research and production use but approach it from opposite directions:}

\begin{figure*}[!t]
\centering
\includegraphics[width=1.5\columnwidth]{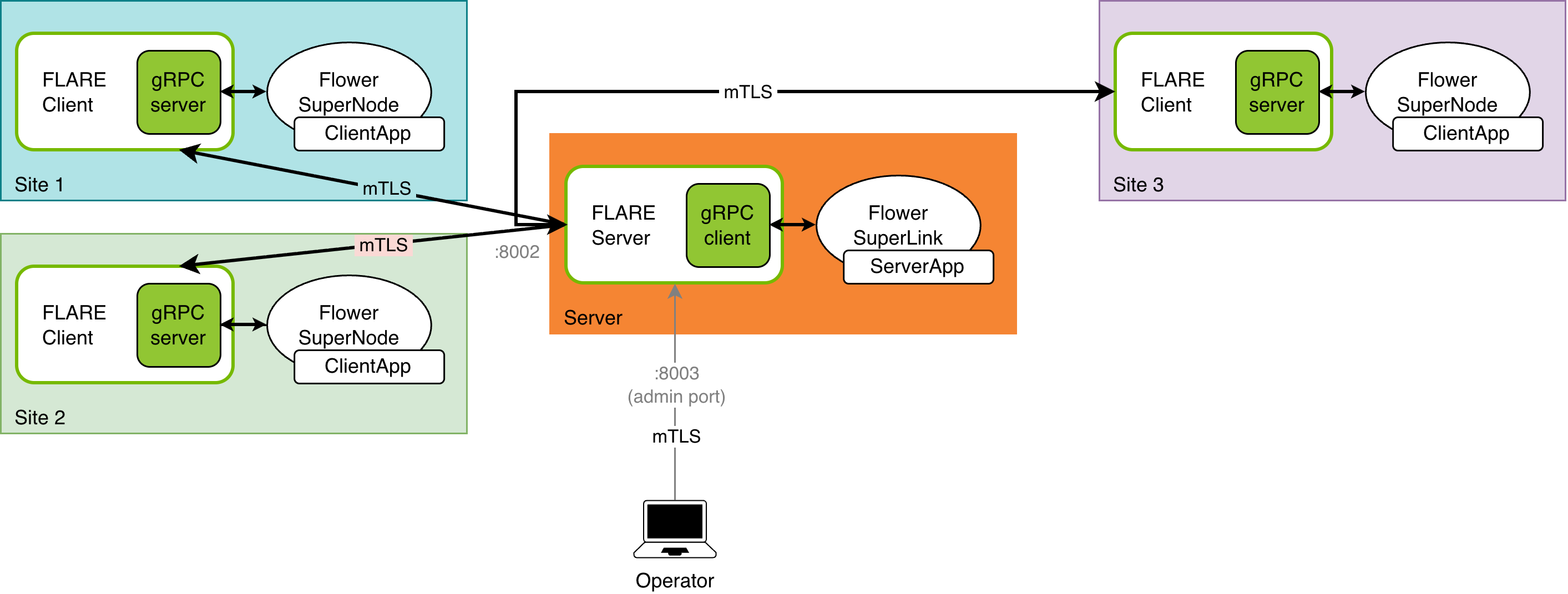}
\caption{\textbf{FLARE-wrapped Flower configuration for three sites (cyan, green, lavender) and a coordination server (black):} Adapted from Figures~2 and~4 of~\cite{roth2024superchargingfederatedlearningflower}. Each Flower \texttt{SuperNode} exchanges unmodified Flower gRPC messages with a local gRPC server inside its site's FLARE client. FLARE relays them over mutual TLS to the default Federated Learning port (\texttt{:8002}) of the FLARE server, and the local gRPC client delivers them to the \texttt{SuperLink}. The operator drives job submission through the default admin port (\texttt{:8003}), also over mTLS. Double-headed arrows denote message flow, with the Client--Server connection initiated by the sites.}
\label{fig:flwr-flare}
\end{figure*}

\begin{enumerate}
    \item{} Flower (\texttt{flwr}) is a lightweight, widely adopted framework with a broad research community. Its architecture uses \texttt{SuperNode} clients that connect to a long-running server, the \texttt{SuperLink}, and it focuses on ease of prototyping;
    \item{} FLARE (\texttt{nvflare}) has a similar client--server foundation but was designed from the outset for production and high-security settings, including air-gapped environments.
\end{enumerate}

The frameworks differ primarily in their default security posture rather than in their ultimate capabilities. FLARE is secure-by-default; its provisioning step issues a startup kit of mutual TLS (mTLS) certificates to every participant, so connections are encrypted, each site is mutually authenticated, and only credentialed nodes may join a run. \textcolor{paul}{Flower supports TLS and node authentication, but these are user-configured.} 
\textcolor{paul}{For deployments over sensitive data, this distinction is significant.}

That said, the two are complementary, and a recent integration~\cite{roth2024superchargingfederatedlearningflower} enables a Flower application to run inside the FLARE runtime (see Figure~\ref{fig:flwr-flare}). 
We adopt the FLARE-wrapped Flower configuration \textcolor{black}{(and the term \emph{site}, as used in that integration's own diagrams, to denote this governance boundary)} throughout because it provides a single deployment path for applications from either framework, without re-engineering the surrounding infrastructure.
\textcolor{paul}{Our deployment uses the FLARE-provisioned mTLS identities, carries Flower traffic inside FLARE's authenticated channels, and restricts inbound connections to the coordination server by source-IP allow-list (Section~\ref{sec:deploy})}.

\section{The Scheduler Mismatch}
\label{sec:mismatch}

General-purpose FL frameworks assume that \textcolor{sadaf}{the server and clients are} reachable throughout the lifetime of a federation. On shared systems, such as those delivered by AIRR, this assumption breaks: jobs enter a queue, receive a bounded wall-clock allocation, and then have their resources reclaimed.
\textcolor{rich}{This mismatch means neither side of the federation can be hosted as an ordinary batch job, and the two cannot be resolved in the same way.}
\textcolor{rich}{In t}his section \textcolor{rich}{we first} establish the deployment configurations in which it arises, then \textcolor{rich}{treat} the coordination server and clients \textcolor{rich}{in turn}.

\subsection{Deployment configurations}
\label{sec:dc}

FL on a shared HPC system appears in three general configurations of increasing isolation, each exercising more of the production federation machinery. In each case, a project is the unit of allocation and access control on its system, granting a fixed set of users a shared filesystem area (what we refer to as a site). In general, a system may host more than one site as separately administered projects, however, in our deployment each system hosts exactly one site with exactly one client.

\begin{enumerate}
    \item \textbf{Single-node simulation.} The coordination server and every client run as separate processes on a single compute node, submitted as one job, within a single site. This is a development configuration that ensures correctness, enables debugging of the FL application, and can simulate a large cross-device federation. The federation never leaves that site, and no network or trust boundary is crossed.
    \item \textbf{Multi-node, single site.} The coordination server and clients occupy separate compute nodes within a single site. Real network connections are now used, and the clients are genuinely distributed, but all nodes share a single administrative boundary and a single data-governance domain. 
    \item \textbf{Cross-silo applications.} Clients run on separate, independently governed HPC systems (e.g., Isambard-AI and Zenith) coordinated by a server that belongs to no site. This is the configuration that motivates FL in the first place, with \textcolor{sadaf}{each site's data behind its own administrative and trust boundary,} \textcolor{rich}{used jointly for training without} \textcolor{sadaf}{being exposed to the other sites}.
\end{enumerate}

Each stage introduces requirements the previous did not. A single-node simulation does not have to authenticate a remote participant or maintain an open connection across a network. \textcolor{rich}{A} single-site deployment \textcolor{rich}{is} distributed \textcolor{rich}{but} still relies on a single scheduler and a single set of user credentials, with \emph{all data} in a single project space. While multiple sites could sit on the same system as individually administered projects, it is only once clients sit on independently administered sites (and, as in our cross-silo deployment, independently administered systems) that authentication across a trust boundary, and the placement of a coordination server that no site owns, become unavoidable. These are precisely the requirements a production federation must satisfy. 

\begin{table*}[!t]
\centering
\caption{\textcolor{black}{Control-Process Placement Regimes and Resource-Hold Behavior.
The inter-job hold is an allocation kept after a federated job completes;
the idle-hold is an allocation kept across the aggregation barrier within a job.}}
\label{tab:regimes}
\begin{tabular}{@{}lllll@{}}
\toprule
Regime & Control process & Inter-job hold & Idle-hold & Realization \\
\midrule
Co-located & in the training allocation & retained & retained & this proof-of-concept \\
Gateway, per-job & decoupled; one allocation per federated job & released & retained & FLARE \texttt{v2.9} Slurm launcher \\
Gateway, per-round & decoupled; one allocation per training round & released & released & under investigation by the FLARE team \\
\bottomrule
\end{tabular}
\end{table*}

\subsection{Server persistence}
\label{sec:sp}
In each \textcolor{rich}{case}, the federation requires a server--client relationship with a persistent coordination server that remains reachable by each client for the federation's lifetime. 

In the first two \textcolor{rich}{cases}, the server can safely run on a compute node within the same allocation as the clients. Because the federation is scheduled together, the server and clients coexist for their lifetime, and when the resources are reclaimed, both are lost together.

In a cross-silo deployment this breaks in three ways. First, each system schedules independently, so a server submitted as a job on one system starts its bounded clock while the clients may still be queuing on theirs, resulting in uncoordinated start times. Second, hosting a server on a compute node would require that node to accept inbound connections, contrary to the systems' network posture, and would commit a whole node to a process that only aggregates and never trains. Third, \textcolor{sadaf}{HPC systems often have bounded wall-clock limits (24 hours on Isambard-AI, for example),} so a server hosted this way may not outlive the client allocations on other systems, and the federation may lose its server mid-run.

The persistently reachable hosts are the login nodes, which serve as shared entry points primarily for job submission. \textcolor{sadaf}{Running a long-lived service or compute-intensive workload degrades them for everyone, and is considered a form of Unacceptable Use under the Acceptable Use Policy\footnote{\url{https://docs.isambard.ac.uk/policies/acceptable_use/\#unacceptable-use}}}.

\subsection{Client placement and idle-hold}
\label{sec:cih}

\textcolor{rich}{The client is affected differently. While the server raises problems for reachability, the client raises problems of placement and cost.} The client control process holds the site credentials and maintains a connection to the coordination server. It drives training on the system it runs on and can admit two deployment regimes: \emph{co-located} or \emph{gateway} (\emph{per-job} or \emph{per-round}), both requiring \emph{a persistent process} but differing in where it can be placed and, in the gateway case, in its driving of the systems's scheduler to acquire and release compute per-round or per-job. Table~\ref{tab:regimes} summarizes all three.

In the \emph{co-located regime}, the client control process is packaged with the training process in a single allocation. The client computes only during local training, sitting idle at the synchronization barrier between rounds while node-hours continue to accrue.
Our proof-of-concept deploys clients in the co-located regime with the \texttt{SuperNode} runtime and the training process packaged in a single Slurm job, holding an allocation for the whole federation that is busy only during training. The idle-hold we measure in Section~\ref{sec:iv-client-cost} is therefore a cost of this deployment choice, made because the gateway regime is not yet deployable with production frameworks. Even then, in the best case, that regime would trade the idle-hold for \emph{per-round queue wait}, and would still require a persistent client control process next to the compute.

This straggler-induced idling is a recognized challenge, and algorithmic solutions have addressed it by adapting local workloads to client throughput (FedCompass~\cite{li2024fedcompassefficientcrosssilofederated}). Reducing how long a fast client waits at the barrier is the algorithmic half of the problem; returning the remaining idle to the scheduler rather than holding it in the federation is the engineering half, and removing the idle-hold needs both. 
Even where a technique reduces the idle fraction, a well-architected deployment should still return compute to the scheduler when it is not training. The hold is, in any case, bounded by the wall-clock limit, so a federation running longer than a single allocation must reacquire through the queue regardless.

\textcolor{black}{The \emph{gateway regime} decouples the control process from the \emph{training allocation}, and comes in two variants that differ in how often compute is released. In the \emph{per-job} variant, the control process is lifted out of the compute allocation and a single training job is submitted per federated job; the allocation returns to the scheduler when the job completes, relieving the inter-job hold, but it is held across the job's training rounds, so the idle-hold persists. This is the variant that will be realized by the first-class Slurm launcher planned for \texttt{v2.9}} (see issue \#4908 on GitHub\footnote{\url{https://github.com/NVIDIA/NVFlare/issues/4908}})\textcolor{black}{, operating at one submission per job. In the \emph{per-round} variant, a job is submitted and released each round, returning the compute nodes to the scheduler at every barrier and removing the idle-hold; the cost is per-round queue wait, which is variable and, on a busy system, potentially long, \textcolor{sadaf}{with resources requested anew each round, rather than once at the start of the job}. Full per-round support is targeted for a future FLARE release.}

\textcolor{black}{The gateway regime can be realized today by persistent middleware. For example, FedQueue~\cite{li2026fedqueuequeueawarefederatedlearning}, a queue-aware FL protocol, submits each round's training job from a service that outlives every allocation.
Unlike the coordination server, this service cannot rely on neutral infrastructure. It holds site credentials, drives training on the system it runs on, and maintains federation state, so it must remain within the trust boundary. Within that boundary, a compute node satisfies the placement requirement but not the persistence requirement; its allocation is as bounded as any other job's, and the login nodes are excluded for the reasons given in Section~\ref{sec:sp}. The client control process needs a host that is in-boundary, persistent, outbound-reachable, and sanctioned, a set of conditions the batch environment cannot provide at once.} 

\begin{figure*}[t]
\centering
\includegraphics[width=1.55\columnwidth]{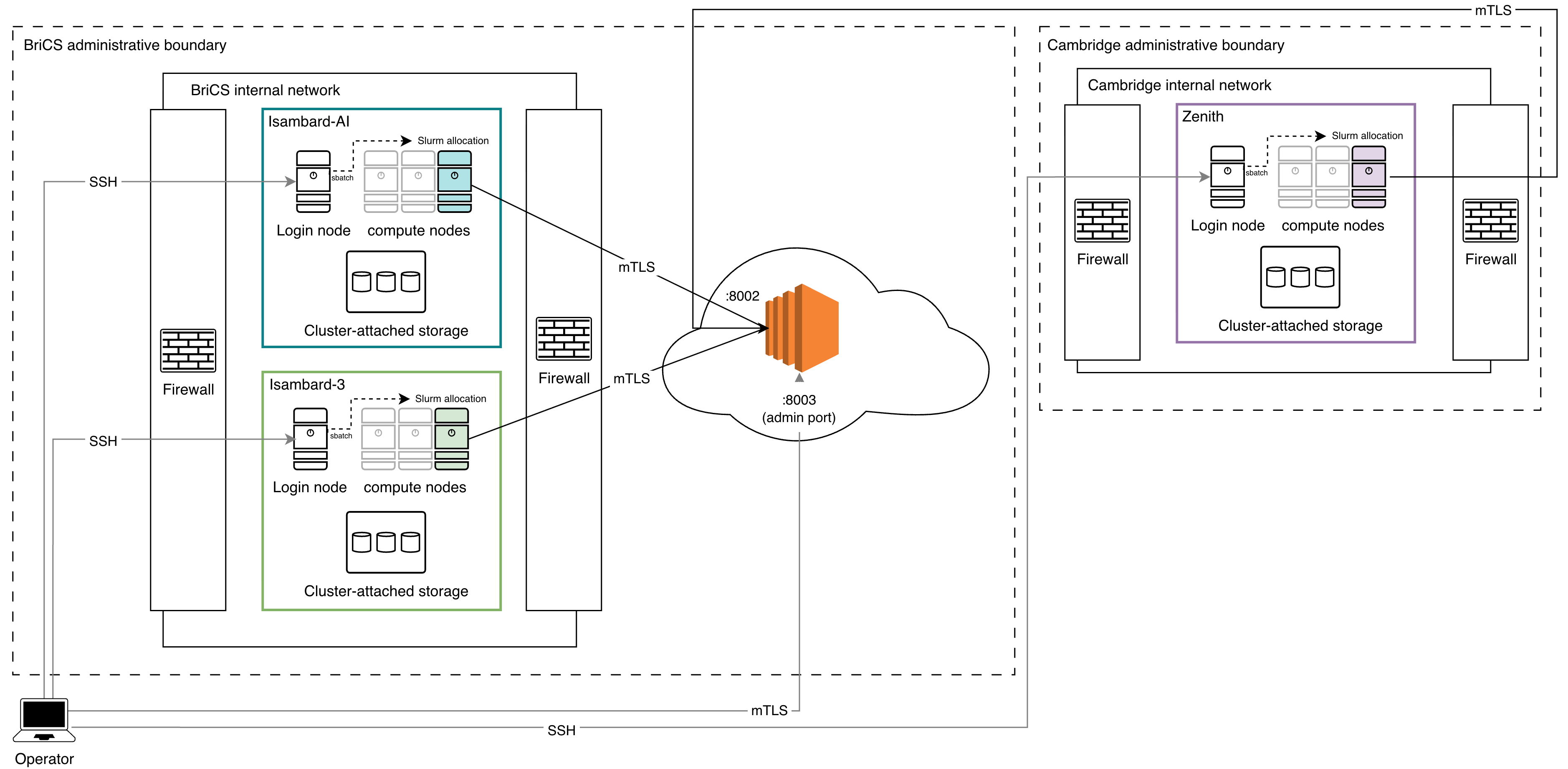}
\caption{\textcolor{black}{\textbf{Cross-silo deployment of the FLARE-wrapped configuration across three systems (Isambard-AI, Isambard 3, and Zenith)}: the setup models the real federation across trust boundaries and systems. Each site is a highlighted compute node within (cyan, green, lavender) hosting the site from Figure~\ref{fig:flwr-flare}. The jobs are launched by \texttt{sbatch} (dashed arrow) from the login node and held within a Slurm allocation for the lifetime of the federation. The EC2 instance (black) hosts the server assembly from Figure~\ref{fig:flwr-flare}, and the solid arrows denote connection initiation, with each connection to the server \emph{initiated outbound} from a site (black) or by the operator (grey), over mutual TLS. No inbound connection crosses a site boundary.}}
\label{fig:iso-hpc}
\end{figure*}

\begin{figure*}[ht]
\centering
\includegraphics[width=1.55\columnwidth]{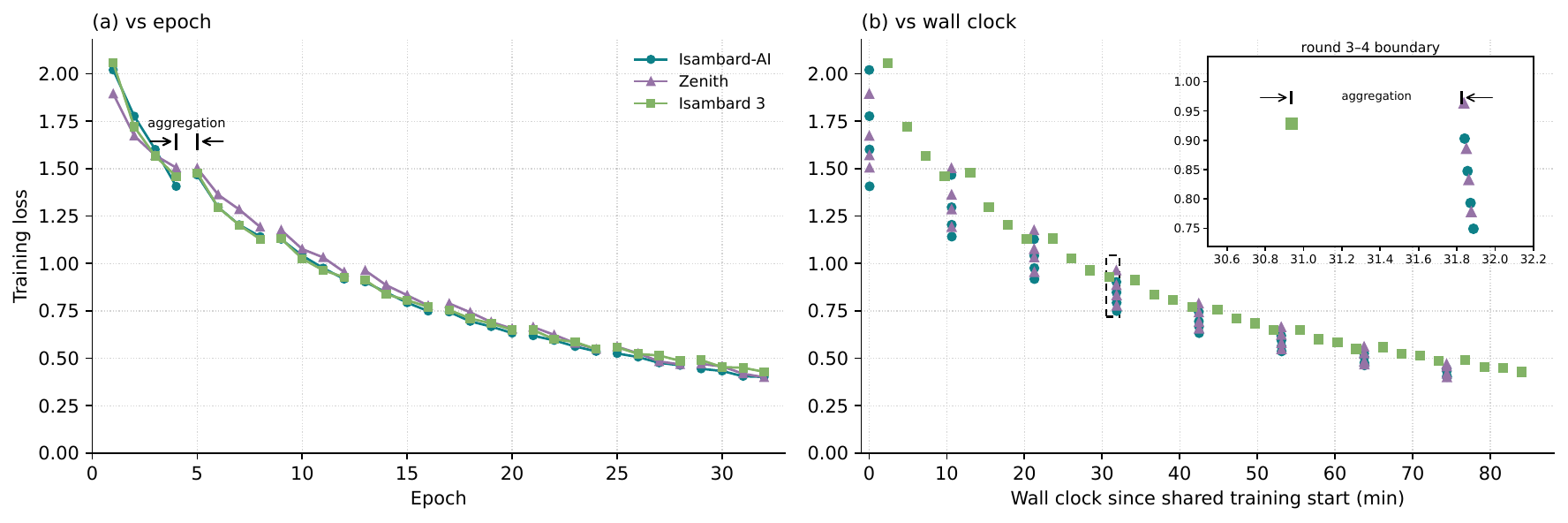}
\caption{\textbf{\textcolor{rich}{Federated t}raining loss \textcolor{rich}{for a Moderate CNN~\cite{Wang2020Federated} on CIFAR-10~\cite{krizhevsky2009learning}}, indexed by epoch (a) and by wall-clock time (b).} Indexed by epoch (a), the three participants track one another closely, converging at essentially the same rate, with solid segments joining successive local epochs within a round; indexed by wall-clock time (b), each participant trains at its own rate, reporting the time of each completed epoch as a point on the graph. Isambard-AI and Zenith complete each round's four local epochs in a few seconds (appearing as a vertical cluster of points) and then hold idle while the CPU-only node (Isambard 3) works through the same four epochs over roughly nine minutes (green squares). The federation only aggregates and advances to the next round when its slowest participant (Isambard 3) finishes.}
\label{fig:training-epoch}
\end{figure*}

\subsection{Summary}
\label{sec:sum}

\textcolor{sadaf}{FL and HPC have mismatched persistence expectations. FL assumes the server and the client control process persist for the federation's lifetime, while HPC provides only bounded allocations.}

\textcolor{sadaf}{The server sees only model updates and control metadata, so it can sit on neutral infrastructure if the federation judges that exposure acceptable; Section~\ref{sec:architecture} demonstrates this with a coordination server on an external EC2 instance.}

\textcolor{sadaf}{The client control process has no such exit. It drives training (and in the gateway regime, submits jobs to the system's scheduler), and is bound to the site's data, so it acts with the site's authority and must stay inside the trust boundary. What it can choose is the deployment regime: either co-located, holding the allocation for the whole federation and paying the idle cost (idle-hold and aggregation time); or gateway, releasing it between rounds (or jobs, depending on the variant) and paying queue wait on each reacquisition (Section~\ref{sec:iv-client-cost}). Either way, a persistent client control process is required (Section~\ref{sec:five}).}

\section{\textcolor{paul}{Cross-Silo} Proof-of-Concept}
\label{sec:architecture}

\textcolor{sadaf}{We assess the feasibility of running unmodified production FL frameworks on batch-scheduled HPC with a proof-of-concept (PoC) implementation. It is not intended as a performance benchmark, but (1) demonstrates cross-silo federation across three independently administered systems at two institutions through a single unmodified deployment path; (2) resolves the coordination server's persistence requirement by placing it outside every participating site; and (3) exposes the facility-level question in running form, holding the client control process in a batch allocation and measuring what that placement costs.}

\subsection{Deployment}
\label{sec:deploy}

We deploy our PoC across Isambard-AI and Zenith (as part of the national AIRR) and Isambard 3, a Tier-2 HPC facility. This is the cross-silo configuration in Section~\ref{sec:dc}, where each system is independently administered with its own network, scheduler, and administrative boundary.

The \textcolor{sadaf}{selection} is deliberately heterogeneous, with an Isambard-AI compute node comprising four NVIDIA GH200 Grace-Hopper Superchips, each pairing a Grace CPU with an H100 GPU; a Zenith node comprising eight AMD Instinct MI355X GPUs; \textcolor{sadaf}{and a CPU-only node on Isambard 3 (built on the NVIDIA Grace CPU Superchip, which combines two Grace CPUs for 144 Arm Neoverse V2 cores, no GPU)~\cite{mcintoshsmith2024isambardaileadershipclasssupercomputer}}. This is in line with expectations in FL, which assumes no common hardware across participants.

The coordination server, a \texttt{SuperLink}, is hosted outside the compute and login nodes of all three systems, on a low-cost virtual machine (an AWS EC2 \textcolor{black}{\texttt{t2.small}} instance in the London [\texttt{eu-west-2}] region). Although administered by BriCS, it sits on infrastructure outside the trust boundary of all three sites, including the two that BriCS itself operates.\footnote{\textbf{In a federation over sensitive data, who operates the coordination server, and who may access the updates it receives, is a decision for the participating sites, and is not constrained to BriCS-operated AWS instances.}}
The coordination server is a natural fit for such placement, \textcolor{sadaf}{as it can persist as long as the federation requires and performs only aggregation}.

Figure~\ref{fig:iso-hpc} shows the topology, with the three sites, each hosting a \texttt{SuperNode} \textcolor{sadaf}{(a GH200 node on Isambard-AI, a Grace-Grace node on Isambard 3, and an MI355X node on Zenith)} opening an outbound connection to the \texttt{SuperLink} on the AWS EC2 instance.  
No inbound port is opened on the HPC side, consistent with each participating system’s network posture.

The VM's security group restricts inbound traffic to the known egress addresses of the three participating systems, limiting access to the coordination server to the federation's own participants rather than the open internet. This network-layer restriction complements the application-layer mTLS (see Section~\ref{sec:background}) so that a connection must both originate from an allow-listed address and present a valid provisioned certificate to join the run, whereas administrative access to the VM is credentialed separately.

While the security group and provisioned certificates govern admission to the federation, access to the sites themselves is governed on a separate identity plane. Users and the operator authenticate to each facility through its own institutional identity provider (IdP), federated under the single sign-on and zero-trust design of~\cite{alam2024}, so each participating institution retains its own IdP.

This resolves the server-side half of the mismatch for this PoC. The server is always-on, requires neither a login node nor a compute node, and no bounded allocation is pressed into hosting it. As established in Section~\ref{sec:mismatch}, the model updates it receives are not information-free; whether their exposure off-site is acceptable is a decision for the federation's threat model. Where governance requires it, the same server could instead run inside a participating institution's own controlled environment, at the cost of provisioning an inbound path to it.

\subsection{Federated Learning}
\label{sec:fl-equivalence} 

Before the cross-silo deployment, we verified that the federation trains correctly in the multi-node, single-site configuration of Section~\ref{sec:dc}, with the server and clients on separate Isambard-AI compute nodes. Equivalence between the FLARE-wrapped and native Flower configurations is established in~\cite{roth2024superchargingfederatedlearningflower}, which licenses the single deployment path adopted throughout.

For the cross-silo deployment, a Moderate CNN~\cite{Wang2020Federated}
was trained on CIFAR-10~\cite{krizhevsky2009learning} using stochastic gradient descent (learning rate $0.01$, momentum $0.9$), a batch size of $32$, four local epochs per round, and eight federation rounds. 
The partitions are independent and identically distributed (IID) by construction, and software versions were pinned at \texttt{nvflare=v2.8} and \texttt{flwr=v1.30}.

Figure~\ref{fig:training-epoch}(a) reports the training loss for each participant as a function of epoch, \textcolor{sadaf}{split by round}. Indexed this way, the three participants track each other closely and train as expected, with the loss decreasing each round.

\subsection{The client-side cost}
\label{sec:iv-client-cost}

The deployment resolves the server half of the mismatch, but the client half persists. A client faces a choice between the two regimes of Section~\ref{sec:cih}, each bearing its own cost.

\subsubsection{\textcolor{paul}{Co-located regime}} 
In the first regime, which our deployment uses, \textcolor{sadaf}{a site's \texttt{SuperNode} runs its \texttt{ClientApp} and training together as a single Slurm job}, holding one allocation for the whole federation. Resources are requested once at \textcolor{sadaf}{job submission} and remain \textcolor{sadaf}{held} for every round; the client never waits to be rescheduled between rounds and need not know in advance how each round will differ. The cost of this deployment is idle-hold. Because the three systems are differently equipped, the training finishes at different times within each round, with the GPU nodes finishing well before the CPU-only node of Isambard 3. The nodes remain connected, but idle.

Figure~\ref{fig:training-epoch}(b) reports the same training loss against wall-clock time. The GPU nodes complete each round's local epochs well ahead of the CPU-only node and then hold at the aggregation barrier. 
Each round takes over $10$ minutes, of which roughly one minute is the aggregation step. Within that round, the participants do \textcolor{sadaf}{the same amount of computation, at very different rates}. The GPU nodes (Isambard-AI and Zenith) complete their four local epochs in a few seconds (appearing as a vertical cluster of points), while the CPU-only Isambard 3 needs about nine minutes. Because aggregation cannot begin until the slowest participant finishes, the GPU nodes then sit idle for roughly nine minutes of every round, about $75$ idle node-minutes each over the eight rounds of Figure~\ref{fig:training-epoch}(b).
The straggler leaves the GPU nodes idle for nearly
$90\%$ of the run. This cost falls on the GPU-equipped participants but lies outside their control, dictated by the slowest participant, not by anything they do, and the co-located regime offers no way to hand a node back before the barrier.

This idle fraction is a property of the testbed, not a general cost of the approach. It is set by the compute mismatch between the slowest and fastest participants. Our deliberate pairing of GPU systems against a CPU-only system drives that mismatch to a near-extreme, placing the fraction near the top of its range. Better-matched participants would see far less; we report it as an existence proof, not a general figure.

\subsubsection{Gateway regime} 

\textcolor{sadaf}{We did not deploy this regime because, at the time of writing, it requires custom middleware~\cite{li2026fedqueuequeueawarefederatedlearning}. As set out in Section~\ref{sec:cih}, the per-job variant (which the first-class Slurm launcher planned for FLARE 
\texttt{v2.9} will provide) decouples the control process and relieves the inter-job hold, but holds the allocation across rounds and so would not remove the idle-hold measured above. Only the per-round variant would remove it. The persistent client control process would acquire the GPU allocation for each round's local training and release it once the local update has been computed, before the straggler-wait and aggregation step. The node would then be charged only for the few seconds it computes, at the cost of a per-round queue wait on reacquisition. Either way, as Section~\ref{sec:cih} established, the regime relocates rather than removes the persistence requirement.}

\subsection{Summary}

The server half is resolved for this deployment. The coordination server 
runs outside every participating site with no bounded allocation pressed into hosting it, sufficient for a single federation under a threat model that accepts that placement. As Section~\ref{sec:five} discusses, an external VM neither scales to many federations nor suits institutions that provision infrastructure differently; the general resolution, for the server and the client control process alike, is the platform developed there. This leaves the client half \textcolor{sadaf}{open}, but neither is fully at home until that platform exists.
Both regimes reduce to the placement asymmetry of Section~\ref{sec:cih}. Whichever cost the client pays, a persistent control process remains inside the trust boundary, and a facility must house it. 

\section{A home for the client control process}
\label{sec:five}
This section is proposed, not evaluated. The measurements of Section~\ref{sec:architecture} cover the external SuperLink and co-located SuperNodes only; neither the gateway regime nor the Kubernetes client placement described here has been deployed.
Section~\ref{sec:architecture} scoped the external-VM placement to this PoC. \textcolor{paul}{The general resolution is to define the host by capability rather than by platform.} The \textcolor{paul_2}{client control process} requires a host that persists independently of any batch allocation, can reach the coordination server outbound, can access the site storage that training requires, and, where training remains a batch job, is authorized to submit and monitor Slurm jobs. These are the capabilities any gateway-regime service requires, whether implemented as middleware~\cite{li2026fedqueuequeueawarefederatedlearning} or as a framework-native launcher; a launcher supplies the software, but the facility must still supply its host. Several realizations satisfy this: a workflow or service node, an adjacent VM within the site perimeter, or, where facility policy permits, an outbound-enabled long-running allocation. 
\textcolor{paul}{The capabilities say what the host must do, and the conditions of Section~\ref{sec:cih} say what it must be to the site:}

\begin{enumerate}
    \item{} \textbf{In-boundary:} operated by the facility, inside the facility's administrative and network perimeter, beside the site's data;
    \item{} \textbf{Persistent:} it outlives any single training allocation, so federation state survives the release and reacquisition of compute;
    \textcolor{paul}{\item{} \textbf{Outbound-reachable}: it can open and hold connections outward to the coordination server, the only network path the federation requires, with no inbound route into the site;}
    \item{} \textbf{Sanctioned:} unlike a daemon on a login node, the host is a supported use of infrastructure the facility provides for exactly this class of workload.
\end{enumerate}

The Kubernetes-as-a-service platform \textcolor{rich}{(Section~\ref{sec:intro})}, \textcolor{sadaf}{which will live alongside the scheduler}, \textcolor{rich}{is our intended solution and is currently under development at BriCS}.
The coordination server and any gateway-regime client control process would then run as Kubernetes services, with the compute-heavy training remaining either as a Slurm-scheduled job, as in the PoC (or as a pod; Section~\ref{sec:v-pod}). The platform absorbs the long-lived control processes, while the scheduler can keep the contended compute it was built to allocate.

\subsection{A fully in-boundary client}
\label{sec:v-pod}

\textcolor{sadaf}{The client control process would run as a long-lived pod, with training in an adjacent, compute pod, scheduled by Kubernetes rather than Slurm. 
Under this architecture, the client would dial outward to the coordination server, leaving the connection model unchanged, but no batch allocation would ever be held or reacquired.}
The persistence requirement \textcolor{paul}{would remain} satisfied, but the \textcolor{paul}{mismatch would be dissolved}, with the process now running where persistence is the platform's purpose.

On the framework side, the Kubernetes job launcher introduced in FLARE \texttt{v2.8} can run each training job as a separate pod, and the Slurm launcher planned for \texttt{v2.9} will submit it to the batch scheduler. Either way, the launcher submits the training job and the facility must still supply the control process's host. Job submission therefore stays scheduler-specific; what the placement design removes is re-engineering the federation's networking, authentication, and control logic around the scheduler.

\subsection{Alignment within the UK TRE community}
\label{sec:v-airr}

In a TRE deployment, the client control process needs a home that is both data-adjacent and inside a governance boundary, exactly the in-boundary, persistent conditions of Section~\ref{sec:cih}. The TRE community is building infrastructure of this shape. 
SATRE~\cite{satre2026} sets a baseline of best-practice capabilities for TREs holding personal data, and K8TRE\cite{Harding_Li_Craddock_Bacon_Silva_Madge_Milligan_Sudhar_Svarovsky_Thomas_2026}, a SATRE-compliant, 
Kubernetes-native TRE, shows that those capabilities are realizable on the cloud-native stack our placement design assumes.
In the DARE UK lineage, FRIDGE\cite{o_reilly_2026_20052251} is a mechanism for extending an existing TRE's governance boundary onto AIRR (a physically separate system, operated with controls sufficient for work carried out within it to be treated as remaining inside the parent TRE's boundary). AIRR-BRIDGE~\cite{airrbridge} is productionizing that model, connecting operational TREs at the Alan Turing Institute and King's College London to AIRR at Dawn and Isambard-AI.

FRIDGE is the natural future placement for the client control process. FL's requirements are modest, with only outbound reachability to the coordination server (the outbound-only pattern of Section~\ref{sec:architecture}) and no inbound route required. 
Because FRIDGE already runs on Kubernetes, the client control process would be a new tenant on infrastructure that has already supported a SATRE-aligned secure data environment, rather than a new deployment model.

\textcolor{sadaf}{A parallel effort within DARE UK} reaches the same outbound-only coordination pattern for Flower by hosting all components on a NetBird 
zero-trust mesh VPN, with each site's firewall permitting only specific addresses. 
One site hosts the \texttt{SuperLink} as the only component accepting inbound connections, restricted to the VPN, while the others connect outbound only. \textcolor{sadaf}{That the same off-scheduler, cloud-native placement and single-inbound-component model also arise under a different transport (a mesh VPN rather than the mTLS-and-security-group approach of Section~\ref{sec:architecture}) suggests the pattern is structural to FL-on-HPC rather than specific to a particular framework or network mechanism.}

\subsection{Generality beyond federated learning}
\label{sec:fivegeneral}

This placement problem is not specific to FL. Any workload that pairs a persistent service with batch-scheduled compute faces the same constraint, and model serving is a clear parallel, as with a high-availability Large Language Model (LLM) inference endpoint.

Existing approaches resolve this within the batch model, either through a Slurm-native service architecture~\cite{doosthosseini_saia_2026} or by brokering inference requests to the scheduler via middleware~\cite{10.1145/3731599.3767346}. However, the same infrastructure-side resolution applies. With a model-serving layer such as KServe running on Kubernetes, both the endpoint and its inference pods would sit entirely off the scheduler, while the scheduler retains the training workload it was built for.

The architecture holds even in the absence of data-governance constraints, making this a generic multi-tenant deployment strategy for persistent services on batch-scheduled systems rather than one specific to FL.

\section{Discussion \& Next Steps}
\label{sec:fin}

We have demonstrated cross-silo FL, using unmodified production FL frameworks, across three independently \textcolor{sadaf}{administered} supercomputers: Isambard-AI and Isambard 3 at BriCS, and Zenith at the University of Cambridge. Three \texttt{SuperNodes} established authenticated connections through FLARE-provisioned certificates to a \texttt{SuperLink} (the coordination server) running outside every participating site, and FL completed as expected. The federation crossed an organizational boundary as well as three system boundaries, and the placement resolved the server half of the mismatch.

The client half remains in the state Section~\ref{sec:iv-client-cost} measured. Holding the allocation \textcolor{sadaf}{pays idle-hold}, releasing it pays queue wait\textcolor{sadaf}{. Either way,} a persistent client control process survives. Its placement, however, is no longer unspecified. Section~\ref{sec:five} states the host as a set of capabilities and four conditions---\emph{in-boundary}, \emph{persistent}, \emph{outbound-reachable}, and \emph{sanctioned}---and one realization is under construction at BriCS. 
What remains open is the interaction between long-lived client processes and scheduler-managed compute, which we intend to investigate using both the first-class (per-job) Slurm launcher planned for FLARE \texttt{v2.9} and the Kubernetes launcher (introduced in the \texttt{v2.8} release) as part of a TRE deployment.

While this work \textcolor{sadaf}{investigates} the infrastructure side of running production FL frameworks on batch-scheduled systems, it treats data access and governance as given. In real-world deployments, particularly those over clinical or other highly sensitive research data, that access is rarely so readily available. Data governance in practice introduces a layer of regulatory constraint over who may initiate a federation and how participating sites are verified. It also raises the question of how privacy-preserving mechanisms (e.g., differential privacy \cite{FLDP} or cryptographic secure aggregation \cite{crypto}) are integrated with a federation and attested to data custodians. By reducing what the coordination server can infer from the updates it receives, these mechanisms make off-site placement acceptable under a wider range of governance regimes.

\textcolor{sadaf}{The deployment of Section~\ref{sec:architecture} was intentionally modest, neither a production-scale demonstration nor a final architecture.
Extending the federation across the edge–cloud–HPC continuum and evaluating production-scale models and datasets are the next steps, beginning with the fully in-boundary Kubernetes client. The first milestone is a single federated round with the client control process in a persistent pod and training submitted to Slurm on Isambard-AI, compared on deployment effort against a middleware-based gateway. At the edge, the outbound-only posture HPC requires is the one such clients already operate under, though source-IP allow-listing would give way to the provisioned mTLS identities as the sole admission control for dynamically addressed clients. AIRRCloud, the emerging cloud arm of AIRR, could provision cloud-based clients that join the federation as the HPC \texttt{SuperNodes} do, but without the batch-scheduler mismatch this paper addresses. The longer-term direction is not to adapt each application individually to the constraints of batch scheduling, but to build infrastructure on which persistent services and scheduler-managed computation coexist---the placement asymmetry we characterize is a first step toward it.}

\section*{Acknowledgments}

The authors acknowledge the use of resources provided by the Isambard-AI National AI Research Resource (AIRR) and the use of resources provided by the Isambard 3 Tier-2 HPC Facility. Isambard-AI is operated by the University of Bristol and is funded by the UK Government's Department for Science, Innovation and Technology (DSIT) via UK Research and Innovation; and the Science and Technology Facilities Council [ST/AIRR/I-A-I/1023]. Isambard 3 is hosted by the University of Bristol and operated by the GW4 Alliance (\url{https://gw4.ac.uk}), and is funded by UK Research and Innovation; and the Engineering and Physical Sciences Research Council [EP/X039137/1]. The authors also acknowledge the use of resources provided by the Zenith supercomputer. Zenith is operated by the University of Cambridge and is funded by the UK Government's Department for Science, Innovation and Technology (DSIT) and UK Research and Innovation (UKRI) as part of the UK Artificial Intelligence Research Resource (AIRR) [ST/Z000890/1], and the National Compute Resource (NCR), and delivered in partnership with Dell Technologies and AMD.

PJW thanks H. Roth and colleagues at NVIDIA for discussions on FLARE that informed this work. HD thanks the EPSRC funded VIVO Hub for Enhanced Independent Living (UKRI142) for their support. 

\bibliographystyle{IEEEtran}
\bibliography{references.bib}

\end{document}